\documentclass[prc,showpacs,superscriptaddress,twocolumn]{revtex4}
\usepackage{graphics,epsfig,colordvi}
\def\be{\begin{eqnarray}}
\def\ee{\end{eqnarray}}
\def\om{\omega}
\def\dsp{\displaystyle}

\def\lsim{\stackrel{\scriptstyle <}{\phantom{}_{\sim}}}
\def\gsim{\stackrel{\scriptstyle >}{\phantom{}_{\sim}}}
\begin{document}
\title{Phi Mesons from a Hadronic Fireball}
\author{Peter Filip}
\affiliation{Max-Planck-Institut f\"ur Physik, D-80805  Munich,
Germany}
\author{Evgeni E. Kolomeitsev}
\affiliation{European Centre for Theoretical Studies in Nuclear
Physics and Related Areas\\ Villa Tambosi, I-38050 Villazzano (TN)
and INFN, G.C. Trento , Italy}
\begin{abstract}
Production of $\phi$ mesons is considered in the course of
heavy-ion collisions at SPS energies. We investigate the
possible difference in momentum distributions of $\phi$ mesons
 measured via their
leptonic ($\mu^+\mu^-$) and  hadronic ($K^+K^-$) decays.
Rescattering of secondary kaons in the dense hadron gas
together with the influence of in-medium kaon potential
can lead to a relative decrease of a $\phi$ yield observed in the
hadronic channel. We analyze  how the in-medium  modifications of
meson properties affect apparent - reconstructed momentum distributions
of $\phi$ mesons.
Quantitative results  are presented for central Pb+Pb collisions at
$E_{\rm beam}=158~{\rm GeV}/A$.
\end{abstract}
\pacs{24.10.Cn,24.10.Pa,25.75.-q,25.75.Dw}
\maketitle
\section{Introduction}
Growing body of experimental information
on fixed-target nucleus-nucleus
collisions from AGS and CERN-SPS accelerators provides
a reliable basis for the systematic investigation of strongly
interacting hadronic matter under extreme conditions.
Particularly, production of particles containing
strange quarks is expected to reflect the reaction
dynamics at the early stage of collisions. In this context,
$\phi$ mesons, particles consisting
mainly of $s\bar{s}$ pairs, are of a special interest. Since
the interaction of $\phi $ mesons with
non-strange hadronic matter is suppressed according to OZI rule,
$\phi $ mesons are expected to decouple easily
from the hadronic fireball.

Several calculations have been done for $\phi$ meson properties
in dense hadronic environment (mainly at higher baryon densities)
\cite{ko,asak,koseibert,haglinphi,songfi,smithhagl,klingl}.
It has been found that modifications of the $\phi$ width and
mass are sensitive to the strangeness content
of the surrounding medium.

Recently, new experimental data from CERN-SPS on
$\phi$ meson production in central Pb+Pb collisions at 158GeV/A
beam energy became available.
Advantage of the rich experimental program at CERN is that it
allows to study the $\phi$ production via different $\phi$
decay channels.
Results of NA49 collaboration ~\cite{na49} are
based on the $\phi$  meson identification via its hadronic decay
$\phi\to K^+K^-$, while NA50 collaboration has recently reported
\cite{na50} about preliminary analysis of $\phi $ mesons
identified via the dileptonic decay channel $ \phi \to
\mu^+\mu^-$.

The purpose of this paper is to show, that the
$\phi $ meson production spectra reconstructed via $K^+K^-$ and
$\mu^+\mu^-$ decay channels can be significantly different
up to the level of present experimental observations.

After introductory considerations in Section~II we derive in
Section~III expressions for momentum distributions of $\phi$
mesons detected via $KK$ and $\mu\mu$ channels. We take into
account rescattering of decay kaons and in-medium modification of
$\phi$ meson properties  discussed in Section~IV. Numerical
results are presented in Sections~V and VI and conclusions are
drawn in Section~VII.

\section{Phi mesons in hadron gas}
For our estimates of the observed $\phi $ meson spectra we
assume a simple two-stage picture of central heavy ion collisions,
which is close to those considered within cascade-transport
\cite{bass98,cass,bravina,sollf,filip}, hydrodynamical
\cite{redlich,toneev,pokrov} and
thermodynamical approaches \cite{pbm,bk,heinz,peitz}.
Initial stage of a collision is characterized  by temperature
close to the QCD  phase transition $T\gsim T_c\approx 170\pm
10$~MeV. Then the system expands up to the point, when numbers of
different kinds of particles freeze in - a chemical
freeze-out. Thermodynamical parameters of this stage could be
obtained by fitting the final total hadron multiplicities. Typical
temperature is found to be $T_{\rm chem}\sim 160\pm 10$~MeV.
During the second stage of the expansion, elastic scatterings
change momentum distributions of hadrons until they cease and
distributions freeze in. The freeze-out temperature can be
extracted from a simultaneous fit to the single-particle
$m_T$-spectra of different particles supplemented by the analysis
of particle correlation data. According to
Refs.~\cite{bk,appel,heinz,peitz} one has $T=T_{\rm therm}\sim
110\pm 30$~MeV.

In our considerations we assume that the fireball created in
heavy-ion collision consists, mainly, of pions, kaons and excited
mesonic resonances in central rapidity region.

Mean free path $\lambda_\phi$
of $\phi$ mesons in hadron gas is
estimated in Ref.~\cite{haglinphi}.
Comparison with mean free paths of pions
and kaons, $\lambda_{\pi,K}$, from Ref.~\cite{haglin,weldon} gives
$\lambda_\pi\lsim\lambda_K<\lambda_\phi$
for temperatures $T_{\rm therm}<T<T_{\rm chem}$.
Hence, one can expect that $\phi$ mesons decouple from the pion-kaon
subsystem at some earlier stage between
the chemical and thermal freeze-out.
Upon this stage  $\phi$ mesons stream freely out from the fireball.
Pions and kaons, on the other hand, may still participate in mutual
secondary interactions up to the stage of total thermal freeze-out.
Therefore, if a $\phi$ meson decays inside a fireball via hadronic channel
we have to take into account  possible interaction of its decay products
with surrounding hadronic environment.

As a result of secondary interactions,
the reconstructed invariant mass of a given pair of $\phi$ daughter kaons
falls out from the original $\phi$ meson peak into the region identified
as a combinatorical background.
Therefore, the $\phi$ mesons decaying in
medium, can be partially unrecognized in experimental analysis of
$K^+K^-$ pairs.
Together with negligible final state interaction of secondary
dimuons originating from $\phi\to \mu^+\mu^-$ decays
this behaviour results in the relative suppression of
$\phi$ meson yield observed via hadronic $KK$ channel.
Such mechanism has been quantitatively studied in Ref.~\cite{suny} where
suppression at the level 40--60\% has been obtained from simulation
using RQMD code.

In this paper we discuss effects which may
enhance the suppression of observed $\phi $ mesons identified
via kaon channel. Possible increase
of a $\phi$ meson width in medium will enlarge
the probability of $\phi $ decay inside a fireball enhancing thus
consequences of mechanism studied in Ref.~\cite{suny}.
Alternatively, we consider
a possibility that the kaon decay channel of a $\phi$ meson
becomes kinematically quenched in the medium.
Additionally we argue that substantial relative
difference in properties of $K^+$ and $K^-$ meson in hadronic environment
caused by the isospin asymmetry and/or by the large baryonic
admixture, would also prevent the reconstruction of $\phi$ mesons
decaying into kaon pairs inside the medium.

\section{Distribution of $\phi $ Decay Products}
Let us denote the phase-space distribution of $\phi$ mesons in the
center-of-mass system of two colliding nuclei at freeze-out as
$f_\phi(\vec{x},\vec{p})$. Then the primary momentum distribution
of $\phi$ mesons is given by \be\label{INTEGR} \eta_0(p) =
\intop_{\Sigma} {\rm d}^3 {\sigma}^\mu\,p_\mu\,
f_\phi(\vec{x}_\phi,\vec{p\,}) \,, \ee where integration goes over
the fireball volume within a freeze-out hyper-surface $\Sigma$
(surface normal vector ${\rm d}^3\sigma^\mu$ contracted with a
$\phi$ meson momentum $p^\mu$) \cite{Cooper}. Here we assume the
case of a time-like freeze-out  hyper-surface, ${\rm d}^3
{\sigma}^\mu\,p_\mu>0$, which is relevant for applications below
(for discussions of alternative cases cf.~\cite{slhs}). In the
absence of any in-medium modifications of decay products and final
state rescattering the shape of {\it observed} momentum
distributions $\eta_0(p)\,\Gamma^0_{\mu\mu}/\Gamma_{\rm tot}$ of
muon pairs $(\mu^+\mu^-)_\phi$ and
$\eta_0(p)\,\Gamma^0_{KK}/\Gamma_{\rm tot}$ of kaon pairs
$(K^+K^-)_\phi$ would be the same. Here $\Gamma_{\rm tot}$ is
$\phi$ meson total width and $\Gamma_{KK}$, $\Gamma_{\mu\mu}$ are
the $\phi $ partial decay widths in kaon and muon decay channels.
Accordingly, in the experimental analysis, $\phi$ meson
distribution is reconstructed from momentum distribution of decay
products multiplied by the corresponding inverse branching ratio.

In this section we derive expressions for the apparent
momentum distributions of $\phi$ mesons
reconstructed via kaon and dimuon decay channels
taking into account possible modifications of
meson properties in medium and consequences of
kaon rescattering.

Let us consider $\phi $ meson suffered the last interaction
at position $\vec{x}_\phi$
inside the hadronic fireball.
The probability, that $\phi$ meson lives for a time $t$ is
\be\label{PBX}
{\rm D}_\phi(t)&=&\exp\Big[-\int_0^{t}
\widetilde{\Gamma}_{\rm tot}^*(t')
{\rm d} t' \Big]\,,
\ee
where $\widetilde{\Gamma}_{\rm tot}=\Gamma_{\rm tot}\,m_\phi/E_\phi$
is total width of a moving meson with energy $E_\phi=(m_\phi^2+p^2)^{1/2}$.
Asterisk * denotes in-medium values of the quantities.

After traveling for a given time $t$
with velocity $\vec{v}_\phi=\vec{p}/E_\phi$
the $\phi $ meson decays
at the position $\vec{x}+\vec{v}_\phi\!\cdot t$
with a probability $\Gamma_{KK}^*(t)/\Gamma_{\rm tot}^*(t)$ into two
kaons.
In-medium values of widths $\Gamma^*$ are determined by
the current local temperature and density of the system.
Daughter kaons from $\phi $ meson decay have momenta
$\pm p_{KK} \,\vec{n}_K$ in the
rest frame of the $\phi$ meson. Value of $p_{KK}$
follows from equation $m_\phi=\om^*_K(p_{KK})
+\om^*_{\bar K}(p_{KK})$
where $\om^*_{K}(p)$ and $\om^*_{\bar K}(p)$
are in-medium spectra of kaons and anti-kaons.
 In the  center-of-mass
system of two colliding nuclei (CMS) the kaon momenta are equal to
\be\label{kmoment}
\vec{p\, }_K^{\pm}&=&\frac{\vec{p}_\phi}{2} \pm \delta \vec{p}\,,
\\ \nonumber
\delta\vec{p}&=& p_{KK}\, \left\{
 \vec{n}_K+\vec{n}_\phi\, (\gamma_\phi-1)\, (\vec{n\,}_K\cdot
\vec{n\,}_\phi)\right\}
\,,\ee
where $\gamma_\phi=(1-v_\phi^2)^{-1/2}$.
The unit vector  $\vec{n}_K$  is uniformly distributed in the
$\phi$ rest frame, direction of $\vec{n}_\phi$ is determined
by $\phi$ meson momentum $\vec{p}_\phi$ in CMS.

For a successful identification of $\phi $ mesons in invariant
mass spectrum of observed $K^+K^-$ pairs it is essential that
momenta of daughter kaons do not change while leaving the hadronic
fireball. Here we investigate two mechanisms which may change
momenta of daughter kaons: rescattering in surrounding hadronic
environment and change of momentum due to in-medium $K$ meson
potential.

Probability that secondary kaon and anti-kaon leave
a fireball without rescattering
is determined by their mean free paths  $\lambda_{K,\bar K}$,
the time which kaons need to reach the fireball border
$\tau_{K,\bar K}^\pm$, and their velocities
$\vec{v\,}_{K,\bar K}^\pm=\vec{p\,}_{K}^\pm/\om^*_{K,\bar K}$
(in CMS) as follows:
\be\label{pl}{\rm P}_\lambda(t) &=&
\exp\!\Big[\!-
v_K^+\!\!\!\!\intop_t^{\tau_K^+ + t}\!\!\!\! \frac{{\rm d} t'}
{\lambda_K(t')}
-v_{\bar K}^-\!\!\!\!\intop_t^{\tau_{\bar K}^-+t}\!\!\!\!
\frac{ {\rm d}  t'}{\lambda_{\bar K}(t')}
\Big]\,.\ee
Thus, the probability to register $\phi$ meson in the kaon
channel when it decays in medium can be expressed as:
\be\label{PROBX}
{\rm P}_{\rm I}\cdot(1-{\rm P}_\lambda(t))+{\rm P}_{\rm II}\cdot {\rm P}_\lambda(t)
\,,\ee
where ${\rm P}_{\rm I}$ and ${\rm P}_{\rm II}$ are probabilities to identify a $\phi$ meson
from rescattered (${\rm P}_{\rm I}$) and non-rescattered (${\rm P}_{\rm II}$) kaons.

Single rescattering of one of the kaons changes momentum
of the kaon pair by vector $\Delta \vec{P}_{\rm scat}$,
with $|\Delta \vec{P}_{\rm scat}|\sim p_T$ being a average
thermal momentum of pions. Correspondingly
the invariant mass changes by $\Delta M^2_{\rm rescat}\sim
p_T^2$\,.
At high temperatures ($T\sim m_\pi$) average thermal momentum
of pions is $p_T\sim 400$~MeV and hence a single rescattering
may shift $M_{K^+K^-}$ far from the $\phi $ meson mass. In this case we
put ${\rm P}_{\rm I}=0$.

Neglecting in-medium effects, one takes ${\rm
P}_{\rm II}=1$, assuming that without a hard rescattering
all kaon pairs from $\phi$ decays can be identified. However the
modification of particle properties can provide another
mechanism preventing $\phi$ identification. In medium
kaons feel an effective mean-field potential
inducing a change of their spectra
$\om^*_{K,\bar K}(p)-\sqrt{m_K^2+p^2}$.
Leaving the fireball, kaons have to come back
to their vacuum mass shell.
This changes the invariant mass and momentum
of the pair. Going out of medium, kaon stays on the same
energy level, and its momentum outside the fireball becomes
${\vec{q}}_{\alpha}^\pm=\vec{p}_K^\pm\,
\sqrt{\om^*_\alpha(p_K^\pm,t)^{2}-m_K^2}/p_K^{\pm\,2}$ ($\alpha=K,\bar K$),
where the kaon energy $\om^*_\alpha$ is evaluated at the moment of
$\phi$ decay. As a result, momentum of the kaon pair is changed
by $\Delta \vec{P}_{\rm pot}\approx \Delta
\om_K^{*\,2}\,\vec{p\,}_K^+/p_K^{+\,2}+\Delta \om_{\bar K}^{*\,2}
\vec{p\,}_K^-/p_K^{-\,2}$, where we  put
$\Delta \om_\alpha^{*\,2}\ll p_K^{\pm\,2}$ and
$\Delta \om_\alpha^{*\,2}=\om_\alpha^{*\,2}(0)-m_K^2$.
The invariant mass shift is then given by $\Delta M_{\rm pot}^2=
(\Delta \vec{P}_{\rm pot})^2$.

In the nucleon free, isospin symmetrical meson gas, in-medium
spectra of kaons and anti-kaons are identical $\om_K(p)=\om_{\bar
K}(p)$ and the invariant mass shift reduces to $\Delta M_{\rm
pot}^2\!\!=\!-\frac{\Delta \om_K^{*\,2}}{2\, p_K^{+\, 2}\,
p_K^{-\, 2}} \Big[\frac12\, p_\phi^4+2 \delta p^2\, p_\phi^2 -2\,
(\delta p\cdot p_\phi)^2\Big]$\,. This expression can be
interpolated between the limit cases $p_\phi\gg p_{KK}$ and
$p_\phi\ll p_{KK}$ as  $\Delta M_{\rm pot}^2\approx -\Delta
\om_K^{*\,2}\,\frac{4\,p_\phi^2}{p_\phi^2+6\,p_{KK}^2}$\,.
Therefore in the symmetrical case $\Delta \vec{P}_{\rm pot}$
and $\Delta M_{\rm pot }^2$ vanish for small $p_\phi$.
Moreover in the isospin symmetrical case kaons suffer only a
small mass modification ($\lsim$30~MeV for $T\sim m_\pi$). Hence,
the release of kaons from a small potential well formed inside a
fireball does not affect the momentum and invariant mass of a pair
strongly enough to prevent a  $\phi$ meson  reconstruction. For
this case we take ${\rm P}_{\rm    II}\approx 1$.

In the case of rather strong isospin asymmetry and/or significant
baryonic admixture $K^+$ and $K^-$ mesons can have quite different
spectra in medium~\cite{kmed}. This may lead to a wide spread of the kaon
pair invariant masses, making the reconstruction of $\phi$ mesons
decaying inside fireball impossible (${\rm P}_{\rm II}\rightarrow
0$). Note, that in this case our results are insensitive to
the details of kaon propagation in medium. They depend only on the
total $\phi$ meson width via function $D_\phi(t)$.

Finally the probability that a $\phi$ meson created at position
$x_\phi$  will be detected via the kaon channel is:
\be\nonumber
&&\intop_0^\infty {\rm d} t\,\widetilde{\Gamma}^*_{KK}(t)
\, {\rm D}_\phi(t)\, {\rm P}_\lambda(t)\, {\rm P}_{\rm II}
\\ \label{kaonprob}
&&=\frac{\widetilde{\Gamma}_{KK}^0}{\widetilde{\Gamma}_{\rm tot}^0}
{\rm
D}_\phi(\tau_\phi)+\intop_0^{\tau_\phi} {\rm d} t\,\widetilde{\Gamma}^*_{KK}(t)
\, {\rm D}_\phi(t)\, {\rm P}_\lambda(t)\, {\rm P}_{\rm II}
\,,\ee
where  $\tau_\phi$ is the time of flight of a $\phi$ meson
through the fireball, for $t>\tau_\phi$ we have
${\rm P}_\lambda=1={\rm P}_{\rm II}$ and ${\rm D}_\phi(t)={\rm D}_\phi(\tau_\phi)
\cdot \exp(-\Gamma_{\rm tot}\,[t-\tau_\phi])$.
Momentum distribution of the kaon pairs from $\phi$
decays is then obtained by averaging expression (\ref{kaonprob})
over all $\vec{n}_K$ directions in the $\phi$ meson rest frame, and
integrating over the fireball volume with a primary $\phi$
distribution - see Eq.(\ref{INTEGR}). Then, multiplying by the inverse
branching ratio we express "reconstructed"
$\phi$ distribution in kaon channel as:
\be  \nonumber \eta_K(p)
&=&\Big\langle{\rm D}_\phi(\tau_\phi)
+\frac{\widetilde{\Gamma}_{\rm tot}^0} {\widetilde{\Gamma}_{KK}^0}
\!\! \intop_0^{\dsp \tau_\phi}\! {\rm d}t {\rm D}_\phi(t)
\widetilde{\Gamma}_{KK}^*(t)\overline{{\rm P}_\lambda(t)}\, {\rm
P}_{\rm II} \Big\rangle \,,
\\  \label{ratrescat}
&&\overline{{\rm P}_\lambda(t)}
=\intop\frac{{\rm d}\Omega_{\vec{n}_K}}{4\, \pi}
{\rm P}_\lambda(t)\,.
\ee
Here and the brackets denote integration:
\be\nonumber
\Big\langle \dots \Big\rangle =
\intop_{\Sigma}
{\rm d}^3 {\sigma}^\mu\,p_\mu\, f_\phi(\vec{x}_\phi,\vec{p\,})
\Big(\dots\Big)\,.
\ee
Note that according to this definition $\eta_0(p)=\langle 1 \rangle$.

Consideration of the di-muon pair momentum distribution
is more straightforward.
With a change of total $\phi$ width in medium
the branching ratio of the $\phi\to \mu\mu$ decay
$\Gamma_{\mu\mu}/\Gamma_{\rm tot}^*$
changes too. Therefore, the apparent $\phi$ distribution
reconstructed in the muonic channel is:
\be \label{ratmu}
\eta_\mu(p) &=&
\Big\langle{\rm D}_\phi(\tau_\phi)+\widetilde{\Gamma}_{\rm
tot}^0\,
 \intop_0^{\dsp \tau_\phi}
{\rm d}t\, {\rm D}_\phi(t)
\Big\rangle\,.
\ee
Note that for $\Gamma_{\rm tot}^*=\Gamma_{\rm tot}^0$ we have
$\eta_\mu\equiv \eta_0$\,, for $\Gamma_{\rm
tot}^*>\Gamma_{\rm tot}^0$ we have $\eta_\mu<\eta_0$
and if $\Gamma_{\rm tot}^*<\Gamma_{\rm tot}^0$
then $\eta_\mu > \eta_0$.

In our numerical estimates we will consider ratio of the reconstructed
$\phi $ meson distributions Eq.(\ref{ratrescat},\ref{ratmu})
which does not depend on the overall normalization:
\be\label{eta}
\mathcal{R}(p)=\eta_K(p)/\eta_\mu(p)\,.
\ee

For a direct comparison with experimental results
on $m_T$ distributions of identified $\phi $ mesons,
one has to integrate over the rapidity interval accessible
to the experiments, utilizing
$p=\sqrt{m_T^2\,\cosh^2 y-m_\phi^2}$. Dependence of
the ratio (\ref{eta}) on $m_T$ reads
\be\label{etamt}
\mathcal{R}(m_T)=<\eta_K(p)>_y/<\eta_\mu(p)>_y\,.
\ee
We also define ratio of the apparent and primary $\phi$ momentum
distributions for $K^+K^-$ and $\mu^+\mu^-$ decay channels as
a function of $m_T$:
\be\nonumber
\mathcal{R}_{K,\mu}(m_T)=<\eta_{K,\mu}(p)>_y/<\eta_0(p) >_y\,
\ee

\section{$\Phi$ decays in medium}
Main hadronic decay channels of $\phi$ meson are
$\phi\to K\bar{K}$ and $\phi\to \rho\pi$.
Let us now consider the change of a $\phi$ decay width in
a hot meson gas due to the
modification of kaon, pion and $\rho$-meson properties.
In-medium properties of pions can be effectively
incorporated by a small
mass shift $m_\pi^*=m_\pi+\delta m_\pi$ with
$\delta m_\pi<<m_\pi$. Similar behaviour is expected for kaons.

Spectral function of $\rho$ meson in medium was extensively
investigated in the context of  di-lepton production in heavy-ion
collisions at SPS energies \cite{rhoN,rhoPi,brown,rhoPigas}.
In baryonic matter,
coupling of  $\rho$ mesons to resonance--nucleon-hole
modes~\cite{rhoN} together with the modification of pions~\cite{rhoPi}
play dominant role. The
$\rho$ meson becomes very broad in medium and
its spectral density strength is driven effectively to lower energies in analogy
to Brown-Rho-scaling picture~\cite{brown}.
In purely mesonic systems the $\rho$ mass is
found to be almost independent of the temperature due to
cancellation of $\pi-\pi$- and $\pi-a_1$-loop
contributions~\cite{rhoPigas}. The $\rho$ width, on the other hand,
is expected to increase in meson gas considerably e.g. by 80~MeV at $T=150$~MeV
and by 160~MeV at $T=180$~MeV~\cite{rapp}.

The partial width of the $\phi\to K\bar K$ decay in medium
depends on the kaon mass $m_K^*$ as:
\be\label{gkk}{\Gamma}_{KK}^*\!=\!
\Gamma_{KK}^0\, p_{\rm cm}^3(m_\phi^2,m_K^*,m_K^*)/
p_{\rm cm}^3(m_\phi^2,m_K,m_K)\,,
\ee
where $p_{\rm cm}(s,m_1,m_2)$ is a kaon momentum in the rest frame
of $\phi $ meson decay,
obeying the equation:  $\sqrt{s}=\sqrt{m_1^2+p_{\rm cm}^2}+
\sqrt{m_2^2+p_{\rm cm}^2}$\,.
Assuming isospin symmetry we have
$m_K=(m_{K^+}+m_{K^0})/2=495.6$~MeV.
Then vacuum  width is equal to $\Gamma_{KK}^0=\{\Gamma_{K^+
K^-}+\Gamma_{K^0 \overline{K^0}}\}/2=1.84$~MeV.
Note that for $\delta m_K>0$ the
in-medium width $\Gamma_{KK}^*$ decreases fast and vanishes for
$\delta m_K\approx 14$~MeV. The
second hadronic decay channel  $\phi\to \rho\pi$  is effected by the
decrease of the $\rho$ meson mass,
increase of the $\rho$ meson width and the Bose-Einstein
enhancement factor for pions.
For the $\phi$ meson at rest we write
\be\label{grho}{\Gamma}_{\rho\pi}^*
=\Gamma^0_{\rho\pi}\,{\kappa_{\rho\pi}(m_\rho^*,\Gamma_\rho^*,T)}/
{\kappa_{\rho\pi}(m_\rho^0,\Gamma_\rho^0,0)}\,,
\ee
and
\be\label{Grho}
\kappa_{\rho\pi}(m_\rho,\Gamma,T) &=& \intop_{2\,m_\pi}^{m_\phi-m_\pi}
\frac{{d}\om}{\pi}\,[(\om-m_\phi)^2-m_\pi^2]^{\frac32}\,
\\ \nonumber
&\times&
\frac{(1+n_\pi(\om))\,\om\,\Gamma/(2\,m_\phi)}
{(\om-E_\rho(m_\rho))^2+\om^2\,
\Gamma^2/(4\,m_\phi^2)}\,,
\ee
where $m_\rho^*$ and $\Gamma_\rho^*$ are in-medium $\rho$ meson
mass and  width, respectively. Vacuum values are $m_\rho^0=770$~MeV,
$\Gamma_\rho^0=150$~MeV, and $\Gamma_{\rho\pi}^0=0.75$~MeV.
We use the constant width approximation for the $\rho$ meson
spectral density and denote
$E_\rho(m_\rho)=(m_\phi^2+m_\rho^2-m_\pi^2)/2\,m_\phi$.
The Bose-Einstein distribution of pions with
temperature $T$ is $n_\pi(\om)$.
In the zero-width limit we obviously have
$\kappa_{\rho\pi}(\Gamma\to
0)=p_{\rm cm}^3(m_\phi^2,m_\rho,m_\pi)\,(1+n_\pi(m_\phi-E_\rho))$\,.
Numerical evaluation of Eq.~(\ref{Grho}) leads  to the
approximated relation
$\Gamma_{\rho\pi}^*\approx
\Gamma_{\rho\pi}^0\,\Big(1-0.91\frac{\delta m_\rho}{100~{\rm MeV}}+0.25
\,\frac{\delta \Gamma_\rho}{100~{\rm MeV}}+0.07\,[\frac{T}{100~{\rm
MeV}}-1]\Big)
$ valid for $\delta m_\rho=m_\rho^*-m_\rho\lsim 200$~MeV,
$\delta \Gamma_\rho=\Gamma_\rho^*-\Gamma_\rho^0\lsim 200$~MeV, and
$100~{\rm Mev}<T\lsim 200~{\rm MeV}$. The shift of
a pion mass produces a minor effect and it is, therefore, neglected
here. Finally, the  total width of $\phi$ is given by
$\Gamma_{\rm tot}^*=2\, \Gamma^*_{KK}+\Gamma_{\rho\pi}^*$\,.
Here and below we do not consider the $\phi$ meson
mass shift, which is small and cancels as soon as we consider the
ratio of the momentum spectra, cf. Eq.(\ref{eta},\ref{etamt}).

For completeness we remark that the
di-lepton decay channel $\phi\to l^+\,l^-$ also
suffers a modification in medium, because
the vector-meson--photon coupling  is suppressed by  meson
fluctuations~\cite{suhong,song}. For the $\phi-\gamma$ coupling
the suppression factor is determined only by kaon
fluctuations $\chi_{\phi\gamma}=(1-2\, \langle|K|^2\rangle_T/f_\pi^2)$.
Here $\langle|K|^2\rangle_T=
\int {\rm d}^3 k \,\exp(-\om_K(k)/T)[(2\,\pi)^3\,2\,\om_K(k)]^{-1}\,$,
$\om_K(k)=\sqrt{m_K^2+k^2}$ and $f_\pi=93$~MeV is the pion decay constant.
Because of the large kaon mass this correction is negligibly small:
$\langle|K|^2\rangle/f_\pi^2\sim 1\%$\,.

Finally we, specify, how the in-medium properties of kaons and
$\rho$ mesons relax to their vacuum values during the fireball
expansion. Assuming  that $\delta
m_{K,\rho}$ and $\delta \Gamma_{\rho}$ are 
proportional
to the density of the system, we have
\be\nonumber
\delta m_{K,\rho}(t) =
\delta m_{K,\rho}^0\,\frac{R_0^3}{R^3(t)}\quad , \quad
\delta \Gamma_{\rho}(t) =
\delta \Gamma_{\rho}^0\,\frac{R_0^3}{R^3(t)}
\,,\ee
where $\delta m_{K,\rho}^0$ and $\delta\Gamma_\rho^0$ are
input parameters.

Before finishing this section we would like to point out that the estimations
done here are valid only for  an almost baryon free fireball.
In the presence of baryons the calculation of a phi
self-energy becomes more elaborated, cf. Ref.~\cite{klingl}.
However, in this case the spectra of kaons
and anti-kaons differ dramatically, inhibiting thereby the $\phi$
reconstruction even more. It corresponds to the case ${\rm P}_{\rm II}=0$
when the reconstructed $\phi$ distributions (\ref{ratrescat},\ref{ratmu})
are the functions of a total width only. To  simulate effectively
the in-medium modification of $\Gamma_{\rm tot}^*$ we will use  eq.~(\ref{gkk})
and vary the kaon mass.

\section{Space-time evolution of the fireball}
After the considerations above let us now specify the
model of the fireball expansion, which we will
 use in our numerical calculations.
Assume a simple homogeneous spherical fireball with the constant
density and temperature profiles. The $\phi$ meson momentum
distribution
\be\label{dist}
f_\phi(\vec{x},\vec{p})=\exp\Big[-\frac{E_\phi
-\vec{p}\cdot \vec{u\,}(\vec{x\,})}
{T_0\,\sqrt{1-u^2(\vec{x\,})}}\Big]
\,,\ee
is determined by the temperature $T_0$, flow velocity profile
$\vec{u}(\vec{x})=v_f\,\vec{x}/R_0$ and radius $R_0$.
Time of flight of a $\phi$ meson and its daughter kaons through the medium
contained in Eqs.~(\ref{pl}) and (\ref{ratrescat}) can be expressed as
$ \tau_\phi =  \tau_R(\vec{v\,}_\phi,\vec{x}_\phi)$ and
$\tau_K^\pm=\tau_{R+v_f\,t}(\vec{v\, }^\pm_K,\vec{x\, }_\phi+
\vec{v}_\phi\,t)$\,, where $\tau_R(\vec{v},\vec{x})$
stands  for a time, during which
a particle with velocity $\vec{v}$
passes a distance from position  $\vec x$
to the border of a sphere with the radius $R$. Since the fireball is
expanding  with radial velocity $v_f$,
this time satisfies the equation
$(\vec{x}+\vec{v}\,\tau)^2=(R+v_f\,\tau)^2$. This implies:
\be \nonumber
\tau_R(\vec{v},\vec{x}\,) &=&
\Big(
\sqrt{ \left(\vec{v\,}\cdot
\vec{x\,}-v_f\,R\right)^2+
\left(R^2-\vec{x\,}^2\right)\,
\left(\vec{v\,}^2-v_f^2\right)}
\\\label{tau}
&-&\left(\vec{v}\cdot
\vec{x\,}-v_f\,R\right)
\Big)\left(\vec{v\,}^2-v_f^2\right)^{-1}\,.
\ee
Solution (\ref{tau}) is valid for $|\vec{x}|<R$ and $|\vec{v}|>v_f$. In
the case $|\vec{v}|<v_f$ we put $\tau=\infty$.

During the expansion
$R(t)=R_0+v_f\,t$\,, fireball density drops as
$\rho(t)=\rho_0\,R_0^3/R^3(t)$ and the temperature decreases as
$T(t)=T_0\,R_0/R(t)$ as expected for relativistic pion gas,
cf. Ref.~\cite{voskjetp}. The kaon mean free
path is $\lambda_K\propto 1/\rho$ and therefore
\be\nonumber
\lambda_K(t) &=& \lambda_K^0 \,{R^3(t)}/{R_0^3}
\,.\ee
To incorporate the freeze-out effect we will assume that as soon as
$T(t)\leq T_{\rm therm}$ kaons become free and
$\lambda_K\rightarrow\infty$ as
well as ${\rm P}_{\rm II}\rightarrow 1 $.
The freeze-out time is then given by
\be\label{taofo}
\tau_{\rm f.o.}=\frac{R_0}{v_f}\, \left(\frac{T_0}{T_{\rm
therm}}-1\right)\,.
\ee
Parameters $R^0$\,, $T^0$\,, $v_f$\,,
and $\lambda_K^0$ serve as input for numerical evaluations below.
The parameter $T_{\rm therm}$ is used for an effective
parameterization of the freeze-out time $\tau_{\rm f.o.}$. It
should not be considered as a true freeze-out temperature,
since our estimation is based on the simplified hydrodynamical
description of a fireball.

The model set up here is a rather crude approximation.
However, the final results are found to be rather
insensitive to the details of hydrodynamical evolution of a
fireball, being determined mainly by the values of $\Gamma_{\rm
tot}^*\, R_0$, $\Gamma_{\rm tot}^*\, \tau_{\rm f.o.}$, and $v_f$.
 We shall vary the input parameters within a broad range
to illustrate different possibilities.

\section{Numerical Estimates}
In this section we perform
numerical evaluation of our expressions for
$\phi $ meson yields reconstructed
via $K^+K^-$ and $\mu^+\mu^-$  channels in central Pb+Pb collisions
at 158GeV/n SPS energy.

First we investigate to what extend rescattering of secondary
kaons enhanced by the in-medium modification of a $\phi$ meson width
can suppress experimentally observed yield of $\phi $ mesons identified
via $K^+K^-$ channel.

We remind that results of Ref.~\cite{suny} give maximal
suppression factor 40\% for the $\phi $ meson observation in
the kaon decay channel.

In our evaluation we use several combinations of input parameters.
Freeze-out temperatures $T_0$ of $\phi $ mesons distributed
according to Eq.(\ref{dist})
vary between $T_{\rm chem}$ and $T_{\rm therm}$:
($i$) $T_0=150$~MeV\,, ($ii$) $T_0=160$~MeV\,,
($iii$) $T_0=170$~MeV\,.
Size of the fireball $R_0$ at the stage of the $\phi$ freeze-out
has to be comparable with $\phi$ meson mean free path $\lambda _\phi $
at given temperature $R_0\approx \alpha_R\lambda_\phi$ with $\alpha_R\sim 1$.
For different temperature parameters above we take according to
Ref.~\cite{haglinphi}:
$\lambda_\phi^{(i)}=13$~fm,
$\lambda_\phi^{(ii)}=10$~fm, and $\lambda_\phi^{(iii)}=7$~fm.

\begin{figure}
\includegraphics[width=5cm,clip=true]{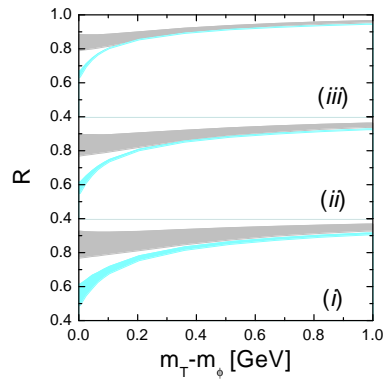}
\caption{Ratio (\protect\ref{etamt}) as a function of $m_T$
calculated for three sets of parameters ($T_0,v_f,R_0)$
without inclusion of in-medium  modifications of kaons and $\rho$
mesons. The upper grey area
corresponds to variations of kaon mean free paths $\lambda_K$ and
freeze-out temperature $T_{\rm therm}$ as described in the text. The
lower grey areas are obtained for the same set of parameters but with the
common expansion velocity $v_f=0.1$\,.}
\label{fig:nomed}
\end{figure}

First we evaluate suppression factor (\ref{etamt})
without any modifications of particle properties in medium, i.e.,
$\delta m_K^0=\delta \Gamma_\rho^0=0$.
In this case we have $\mathcal{R}_\mu(m_T)\equiv 1$ and
$\mathcal{R}(m_T)=\mathcal{R}_K(m_T)$.
Flow velocities corresponding to selected temperatures
$T_0$ are  adjusted to reproduce the slope of $\phi $ meson $m_T$
distribution measured by the NA50 collaboration: $T_{\rm
eff}=218$~MeV: $v_f^{(i)}=0.50$\,, $v_f^{(ii)}=0.46$\,,
$v_f^{(iii)}=0.41$\,.
We take $\alpha_R=1$ and vary the
mean free path of kaons $\lambda _K$  within the interval
$0<\lambda_K^0<\lambda_K(T_0)$, where $\lambda_K(T_0)$ follows from
estimations of
Ref.~\cite{haglin}: $\lambda_K^{(i)}=2$~fm,
$\lambda_K^{(ii)}=1$~fm,
and $\lambda_K^{(iii)}=0.5$~fm. We vary also $T_{\rm therm}$ between
100~MeV and 80~MeV in agreement with analysis~\cite{heinz}.
This translates into the interval of
freeze-out time values $10~{\rm fm}<\tau_{\rm f.o.}<20~{\rm fm}$. All these
variations produce the upper grey areas shown in
Fig.~\ref{fig:nomed}

For all three sets of parameters  $(T_0,v_f,R_0)$
we observe that  $\mathcal{R}(m_T)$ does not
fall below $0.8$ significantly. This is related to the large
expansion velocity of the fireball. In this case
$0.2<\Gamma_{\rm tot}^0\,\tau_{\rm f.o.}<0.4$ and $\phi $ mesons
decay after the thermal freeze-out.
To illustrate this
effect we recalculate ratio $\mathcal{R}(m_T)$  for the same
three cases fixing $v_f=0.1$ what corresponds to
$\Gamma_{\rm tot}^0\,\tau_{\rm f.o.}\sim 1$.
Results obtained ($\mathcal{R}\sim 0.6$) are shown as
lower gray areas in Fig.~\ref{fig:nomed}.

\begin{figure}
\includegraphics[width=5cm,clip=true]{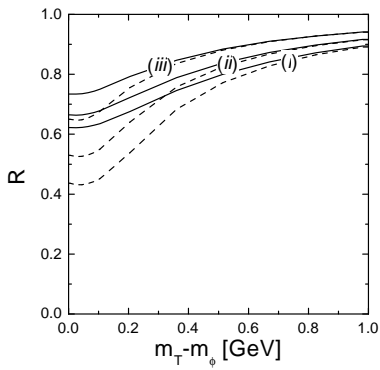}
\caption{Ratio (\protect\ref{etamt}) calculated for cases
$(i)$--$(iii)$ with  $\alpha_R=1.5$, $\lambda _\phi = 13$~fm and
$\lambda_K^0=0.5$~fm. Solid
lines are calculated with $T_{\rm therm}=80$~MeV and dash lines
correspond to $T_{\rm therm}=40$~MeV.
No medium effects are included.}\label{fig:nomedR15}
\end{figure}

To reproduce results of RQMD calculations described in
Ref.~\cite{suny} we take somewhat larger size of a fireball
with $\alpha_R=1.5$, freeze-out temperature
$T_{\rm therm}=80$~MeV and $\lambda_K^0=0.5$~fm.
This corresponds to the lowest
limit allowed by the analysis~\cite{heinz}. The results are shown in
Fig.~\ref{fig:nomedR15} by solid lines. The
limiting scenario considered in \cite{suny}, when the freeze-out
volume is determined by the last kaon interactions, can be
reproduced with $T_{\rm therm}=40$~MeV. This case is shown by dash lines
in Fig.~\ref{fig:nomedR15}.
We take solid lines in Fig.~\ref{fig:nomedR15} as a
reference point for our further investigation of  in-medium effects.

First we consider modifications of $\rho$ meson properties. We choose
$\rho $ mass shift to be $\delta m_\rho^0=-200$~MeV for our three
parameter sets.
The $\rho $ meson width depends on the temperature. Relying on
Ref.~\cite{rapp} we
take: $\delta\Gamma_\rho^{(i)}=80$~MeV,
$\delta\Gamma_\rho^{(ii)}=180$~MeV, and
$\delta\Gamma_\rho^{(iii)}=200$~MeV.
Comparison with Fig.~\ref{fig:nomedR15} (solid lines) shows slight
decrease of ratio $\mathcal{R}(m_T)$ which corresponds to
small increase of  the total $\phi$ meson width by 40\% due to
$\phi\to\rho\pi$ channel.

\begin{figure}
\includegraphics[width=8cm,clip=true]{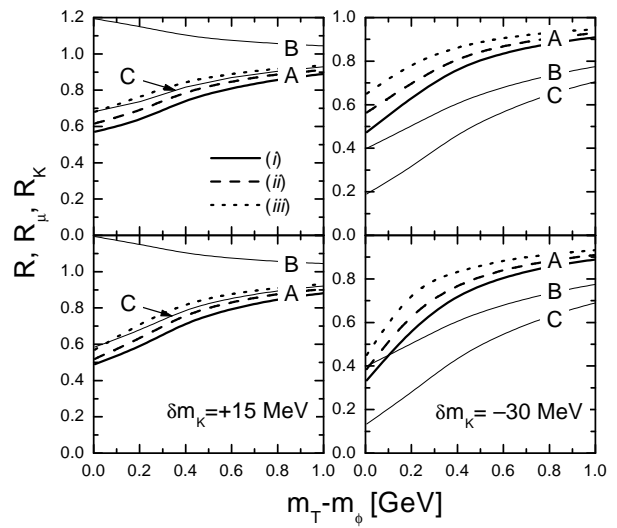}
\caption{Thick lines (label A) show the ratio $\mathcal{R}(m_T)$
calculated for in-medium modification of $\phi$ meson properties.
Results for different parameters sets $(i)$--$(iii)$ are depicted by
solid, dash and dotted lines, respectively.
The left plane corresponds to the case when
the  kaon mass increases in medium $\delta m_K^0=15$~MeV,
whereas the right plane shows results
for a decreasing kaon mass $\delta m_K^0=-30$~MeV.
Thin lines show suppression factors in
the muon $\mathcal{R}_\mu$ (B) and kaon $\mathcal{R}_K$ (C)
channels calculated for parameter set ($i$). Upper plots
are calculated for the
case ${\rm P}_{\rm II}=1$, lower plots correspond to ${\rm P}_{\rm II}=0$.
In all cases  modification of  $\rho$
meson properties in medium is taken into account.}
\label{fig:med}
\end{figure}

Fig.~\ref{fig:med} shows results obtained taking into account
modification of $K$ meson properties in-medium.

First we investigate the case when $\phi\to K\bar{K}$ channel
is closed initially ($\delta m_K^0=15$~MeV)
and it opens only during the fireball expansion.
Results are shown  in the left part of
Fig.~\ref{fig:med}, where the upper plot is calculated for
${\rm P}_{\rm II}=1$ and the lower one corresponds to a strong
suppression of the kaon channel in medium with ${\rm P}_{\rm II}=0$.
Comparing ratios $\mathcal{R}(m_T)$ in Fig.~\ref{fig:nomedR15}
and in Fig.~\ref{fig:med} shown by thick solid line (A)
calculated for parameter set ($i$)
we observe that kinematical quenching of $\phi \rightarrow K^+K^-$ channel
by increasing kaon mass
decreases ratio $\mathcal{R}(m_T)$ very slightly for both values of
${\rm P}_{\rm II}$.
This happens because $\phi$ meson width
becomes  very small in this case and therefore probability of $\phi $
meson decay inside the expanding fireball and consequently also probability
for rescattering of daughter kaons are small.
Compare lines (C), calculated for parameter set ($i$), with a
corresponding line in Fig.~\ref{fig:nomedR15}.
Lines (B) in  Fig.~\ref{fig:med} (left side) show that $\mathcal{R}_\mu$
becomes larger than one for increasing kaon mass.
Since increase of $R_\mu$ by 10\%--20\%
does not change considerably the effective $m_T$ slope of
$\langle \eta _\mu \rangle _y$ distribution, we do not need
to readjust the flow velocity parameter.  This increase of
$R_\mu$ leads at the end to a small decrease of $\mathcal{R}$.

Let us now consider the case when the $\phi$ meson width increases
strongly in hadronic medium due to the increase of $\Gamma_{KK}^*$.
We simulate this effect by decrease of the kaon mass in
medium, which can result e.g. from rescattering
of kaons on pions through $K^*$ and heavier kaonic resonances
\cite{shur}. Here we restrict ourselves to rather conservative modification
of kaon masses $-30$~MeV$ <\delta m_K^0 <0$ which corresponds to
the $\phi$ width $4~{\rm MeV}\lsim \Gamma_{\rm tot}^*\lsim  20$~MeV.
In this case $R_\mu<1$ and at small $m_T-m_\phi$ region
$R_\mu$ can be suppressed up to
40--60\%.
Thus, for a given freeze-out temperature $T_0$ and
total width $\Gamma_{\rm tot}^*$ we readjust flow
velocity $v_f^0$ to reproduce the slope of the $m_T$ distribution
measured in di-muon channel by NA50 \cite{na50}.
For our three sets of parameters ($T_0, R_0$)
we obtain new flow velocities:
($i$) $v_f^0=0.38$\,, ($ii$)
$v_f^0=0.35$\,, ($iii$) $v_f^0=0.28$.

Right part of Fig.~\ref{fig:med} shows our results obtained for
$\delta m_K^0=-30$~MeV.
Corresponding partial width is
$\Gamma_{KK}^*\approx 10$~MeV, and the total width is
$\Gamma_{\rm tot}^*\approx 21$~MeV.
This leads to
$\Gamma_{\rm tot}^*\,R_0\sim\Gamma_{\rm tot}^*\,\tau_{\rm f.o.}\sim
2$,
which provides a strong suppression of $R_K$ as it is shown
for the  parameter set ($i$)
in Fig.~\ref{fig:med} (right side, lines C) . We find
$\mathcal{R}_K(m_T\to m_\phi)\sim 0.2$ for
${\rm P}_{\rm II}=1$ and $\sim 0.15$ for ${\rm P}_{\rm II}=0$.
However, since the ratio  $\mathcal{R}_\mu(m_T)$ also shown in
Fig.~\ref{fig:med} (lines B) is also suppressed
the resulting ratio $\mathcal{R}(m_T)$ remains on the
level $\sim 0.3$ for small $m_T-m_\phi$, provided we put ${\rm P}_{\rm II}=0$
and $\sim 0.5$ for ${\rm P}_{\rm II}=1$.
Taking even larger values of the total decay width
in the most preferable case ($i$)
we obtain
$\mathcal{R}(m_T\to m_\phi)\approx 0.28$
for $\Gamma_{\rm tot}^*=27$~MeV, $v_f^0=0.35$, and
$\mathcal{R}(m_T\to m_\phi)\approx 0.23$ for
$\Gamma_{\rm tot}^*=34$~MeV, $v_f^0=0.32$
(both for ${\rm P}_{\rm II} = 0$).

\begin{figure}
\includegraphics[height=5cm,clip=true]{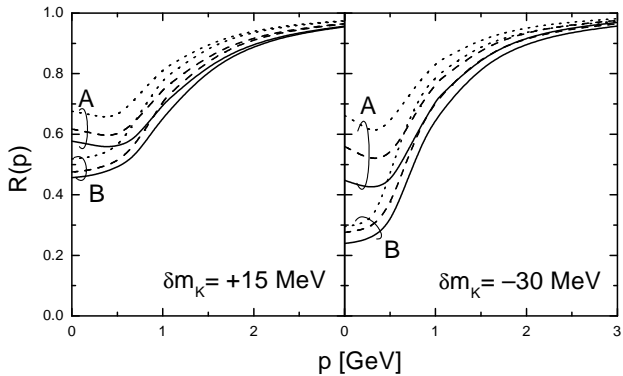}
\caption{
The ratio (\protect\ref{ratrescat}) as a function of $\phi$ meson
momentum. Curve styles correspond to those in
Fig.~\protect\ref{fig:med} with readjusted flow velocities.
Curves A are calculated for ${\rm P}_{\rm I}=1$,
whereas sets B are
calculated with ${\rm P}_{\rm I}=0$.
}\label{fig:pdep}
\end{figure}

It is instructive to investigate ratio (\ref{eta}) as a
function of the $\phi$ momentum. Averaging over rapidity mixes
momenta within a broad interval, e.g., for $m_T=0$ momenta $100\lsim
p\lsim 1000$~MeV what partially washes out the
 final suppression
effect. To illustrate this point we show
ratio $R(p)$ as a function of a $\phi$ momentum in Fig.~\ref{fig:pdep}.
For increasing kaon mass and
correspondingly vanishing $\Gamma_{KK}$ we obtain a decrease of
$R(p)$ at small momenta  by 40--50\% (left plane).
This is again a direct consequence of a rapid fireball
expansion which brings $\Gamma_{KK}(t)$ in the integral
(\ref{ratrescat}) fast to its vacuum value.
In the extreme case for $p\to 0$ and $v_f\to 0$ we get $R(p,v_f)\to 0$.
For decreasing kaon mass the reduction is even stronger $\sim$70\%.
Momentum dependence of $\mathcal{R}$
at small momenta $p\lsim p_f=
m_\phi\,v_f\, \gamma_\phi$ is completely washed out by
rapidity averaging.

\begin{figure}
\includegraphics[width=5cm,clip=true]{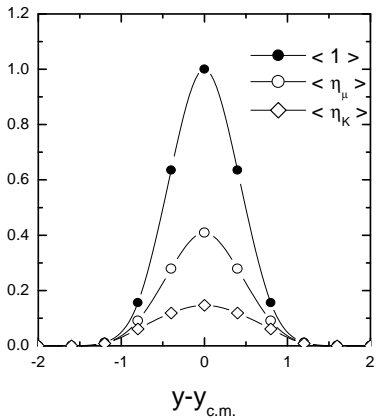}
\caption{
Rapidity distributions $\langle 1 \rangle(y)$ (full circles),
$\eta_\mu(y)$ (open circles)  and
$\eta_K(y)$ (open diamonds) calculated for the parameter set ($i$) with
$\delta m_K^0=-30$~MeV, the flow velocity $v_f^0=0.38$ and
${\rm P}_{\rm I}=0$.
}\label{fig:ydist}
\end{figure}

In Fig.~\ref{fig:ydist}
we compare the original rapidity distribution of
$\phi$ mesons $\eta_0(y)$,
with the distributions which can be reconstructed
via kaonic $\eta_\mu(y)$, and muonic $\eta_K(y)$ decay channels.
Calculations are done for the parameter set ($i$) with $
\delta m_K^0=-30$~MeV, flow velocity $v_f^0=0.38$ and
${\rm P}_{\rm II}=0$. Distributions are normalized to give $\eta_0(y=0)=1$\,.
We observe a considerable broadening of the rapidity
distribution measured in the kaon channel.

\section{Conclusions}
We have studied distributions of $\phi $ mesons
in heavy-ion collisions at SPS energies
reconstructed via hadronic $K^+K^-$ and dilepton $l^+l^-$ decay channels.
The analysis of $\phi $ meson mean free path allows to suppose that $\phi$
mesons decouple from the hadronic system at somewhat earlier stage before
the common breakup of the hadronic fireball. Therefore, kaon pairs
originated from the $\phi$ decays inside a fireball can be
rescattered or absorbed. Such kaon pairs will not contribute to a
$\phi$ meson reconstruction, whereas the leptonic probes can leave a
fireball freely. We derive the expressions (\ref{ratrescat}) and
(\ref{ratmu}) for the apparent momentum distribution of
$\phi$ mesons in kaonic and muonic channels respectively.

Within a simple model of spherically expanding fireball we
investigate dependence of a relative suppression factor of the
hadronic channel with respect to the dileptonic one
on parameters of the system
and on the $\phi$ meson in-medium properties. For a vacuum $\phi$-meson
width $\sim 4$~MeV  the maximal suppression  0.6--0.8 is obtained for the
fireball size and expansion time $R_0\sim \tau_{\rm f.o.}\sim
20$~fm. These values are in agreement with results of RQMD simulations
\cite{suny}. The crucial parameter is the fast expansion of a fireball with
$v_f\sim 0.4$--0.5 corresponding to the $\phi$ freeze-out temperature
range $T_0\sim 150$--170~MeV.

Width of hadronic $\phi$ meson decay channels ($\phi\to K\bar{K}$ and
$\phi\to\pi\rho$) can be modified in medium due to changes of
the meson properties.
We have found  that  increase of the $\pi\rho$ channel width due
to the broadening of $\rho$ meson and decrease of $\rho $ meson mass
leads alone  to a tiny increase of the suppression.

Other
possibility is kinematical quenching of the kaon decay channel, which
we simulate by simultaneous increase of $K^+$ and $K^-$ masses.
Since total width $\Gamma ^*_{\rm tot}$ of $\phi $ meson in medium
becomes small (increase of the $\pi\rho$ channel width is not strong enough)
$\phi$ mesons decay mainly outside the fireball, where vacuum
properties of $\phi $ mesons are restored and rescattering of
daughter kaons is negligible.
Together with relative amplification of the muon decay channel
by 20\% the resulting suppression factor found for a quenched kaon decay channel
was $\sim 0.5$.

The increase of the $\phi$ meson width in medium  provides, on the
other hand,
a mechanism for strong suppression ($\sim 0.15$) of the kaonic detection
channel due to the enhancement of $\phi $ decay probability inside a
fireball increasing thus rescattering of daughter kaons.
However, increase of the $\phi$ total
width reduces simultaneously the branching ratio of
$\phi\to\mu^+\mu^-$ decay and suppresses the spectrum of $\phi $ mesons
reconstructed via $\mu^+\mu^-$ decay channel.
Obtained suppression of the muonic decay channel at the level
$\sim 40$--60\% requires readjustment of the flow velocity to be
compatible with experimental slope of NA50.
Adjusted flow velocity $v_f\sim 0.3$--0.4 for
$\Gamma_{\rm tot}^*\sim 20$~MeV  and $T_0\sim 150$--170~MeV
(compare to $v_f\sim 0.4$--0.5 obtained for a vacuum $\phi$ width)
gives final net relative suppression factor of kaon channel to muon
channel $\sim 0.3$. This value is close to experimental observations
at CERN SPS \cite{na49,na50}.

Strong increase of $\phi $ meson in-medium width can take place
if the kaon mass decreases in medium by 30~MeV.
The mechanism for such kaon mass modification can be
similar to that studied in Ref.~\cite{shur}.

We have found that reconstructed rapidity distributions
of $\phi $ mesons become effectively wider, if in-medium properties
of mesons and rescattering of kaons are taken into account.

Finally we suppose that to improve understanding of
experimental results on $\phi $ meson production in
heavy ion collisions at
CERN SPS \cite{na49,na50} further detailed investigations
taking into account in-medium effects
within transport or hydrodynamical models are necessary.
\begin{acknowledgments}
The authors would like to thank Yu.B.~Ivanov, K.~Redlich, V.D.
Toneev, and D.N.~Voskresensky for discussions. Work of E.K. was
supported by the grant of ECT$^*$, P.F. was partially supported by
Slovak Grant Agency for Sciences Grant No. 2/5085/00.
\end{acknowledgments}


\end{document}